# DNA properties investigated by dynamic force microscopy

L. Nony, R. Boisgard, J.-P. Aimé[1]

CPMOH

UMR 5798 CNRS, Université Bordeaux I

351, cours de la Libération 33405 Talence Cedex

## Abstract

In this work, we show that by varying the experimental conditions, the driving amplitude, a dynamic force microscope allows DNA properties to be selectively imaged. The substrate on which the DNA is fixed is a silica surface grafted with silanes molecules ended with amine groups. Use of small oscillation amplitudes favors the attractive interaction between the tip and the sample, while use of large amplitudes renders the contribution of the attractive interaction negligible. Particularly, at small amplitudes, the images show that the attractive interaction is strongly enhanced along the DNA. This enhancement is found to be amenable with a model considering a narrow strip of randomly oriented dipoles on each side of the molecule. This work should provide new insights on the DNA interaction and conformational changes with localized charges.

---

[1] Corresponding author : jpaime@cribx1.u-bordeaux.fr, 33 (0)5 56 84 89 56

## I- Introduction

Investigating DNA conformation and structure is a long-standing goal in which visualization techniques have a central importance. Therefore, since the beginning of the Scanning Force Microscopy, a lot of experiments have been dedicated to studies of DNA deposited onto a surface [1-5]. Many of them were dedicated to determine the influence of the substrate on the chain conformation [6-8]. It is a daily difficulty to evaluate the influence of the sample preparation, then to be able to extract information of some use for biological purpose. Therefore a wide variety of experiments were done, many of them being focussed on the surface preparation : mica, grafted mica or grafted silica surfaces with molecules ended with amine groups [6, 8-10]. Other attempts aimed at studying the effect of DNA solutions with different ionic strength and pH, particularly the influence of various di-cations was investigated [11-14]. In many cases, experimental strategies are based upon the knowledge accumulated with electron microscopy [15]. The delicate balance to achieve is, on one hand, to prepare a surface being able to stick the DNA and on the other hand, to meet the contradictory requirement of being able to spread the molecule to prevent formation of aggregates.

Beyond this difficulty, the Scanning Force Microscopy (SFM) is a tool that measures the interaction between a nanotip and a surface or an object. Particularly, for soft material, rather than topography, SFM probes the mechanical properties [16]. Among the SFM, the dynamic force microscopy (DFM) has been found as the suitable tool to investigate surface morphology of soft material and then has been widely used to image DNA structure. DFM, namely Tapping mode, was first conceived to reduce the contact area between the tip and the surface, so that the shear forces at the interface are reduced. In addition, recent experimental results and theoretical developments have emphasized that the non-linear behavior of the oscillating nanotip at proximity of a surface provides a very high sensitivity of the force measurement [17-19]. Besides these findings, despite the name of Tapping mode, it was shown and demonstrated that intermittent contact and purely non-contact situations between the tip and the surface can occur [20-22]. When intermittent contact situations take place, the repulsive tip-sample interaction is dominant and the image gives information about the topography and the mechanical properties of DNA. When purely non-contact situations take place, the attractive tip-sample interaction is dominant, and the image gives information about the "chemistry" and the different types of interaction that can happen.

In many cases, one obtains a mixing of the two regimes, so that it is worth discussing the oscillation behavior of the tip-cantilever system as being the result of a dominant repulsive or a dominant attractive regime, rather than to describe the different experimental situations as the result of intermittent contact situations or purely non-contact ones.

As shown below, the two regimes can be straightforwardly discriminated by looking at the oscillation behavior, particularly the phase variation. During an experiment, the different regimes are finely tuned by varying the oscillation amplitude [21]. Qualitatively, the effect of the oscillation amplitude can be understood as follow : let's consider a cantilever that oscillates with an amplitude A at a distance D of the surface (fig.1a), with D > A. For a fixed closest distance between the tip and the surface, $\Delta = D-A$, the time during which the tip is at proximity of the surface is a function of the oscillation amplitude [23]. Use of large amplitudes leads to a shorter residence time than the one obtained at small amplitudes thus to a smaller attractive interaction. A quantitative evaluation is readily deduced from a power law dependence describing the attractive interaction between the tip and the surface. For example, using an effective Van der Waals sphere-plane interaction gives an average attractive force varying as $\Delta^{-3/2}$ from which is derived a strength of the attractive interaction scaling as $A^{-3/2}$ [23-25].

In other words, depending of the regime chosen, one may expect to access to various physical and chemical properties of the DNA molecule at the local scale, then to learn a little bit more about the way the DNA molecule interacts with its surrounding substrate.

Taking advantage of these two possibilities, a DNA molecule (a 2500bp fragment of a linearized plasmid, pSP65) was investigated. The DNA solution was prepared with 10mM $MgCl_2$, 50 mM NaCl and 10 mM trisEDTA (pH 7) with a concentration of 1µg/ml. Then a droplet of 10µl was deposited during 2 minutes on an aminopropyltriethoxysilanes (APTES) grafted silica surface. The surface treatment for the grafting is given in detail elsewhere [26]. The sample was then rinsed with 30µl of deionized water and dried with a filter.

## II- Experimental results

*II-1- Experimental methodology*

The properties of the oscillating tip-cantilever used [27] are the following : the resonance frequency is $\nu_0 = 153319$ Hz, the quality factor is Q = 410, the experiments were performed at u = $\nu/\nu_0$ = 0.9989, corresponding to a phase $\varphi_{free}$ = -45° and an oscillation amplitude $A_{free} = A_0/\sqrt{2} \approx 0.707 \times A_0$ for large tip-surface distances, where $A_0$ is the amplitude at the resonance frequency and $A_{free}$ the oscillation amplitude. The subscript "free" means that the oscillation conditions are measured at a tip-surface distance for which the tip does not interact with the surface, typically distances of ten nanometers or more. The working amplitude, A, e.g. the oscillation amplitude used as a set point to control the vertical location of the surface with the feedback loop, is always set at A = $0.67 \times A_0$, whatever the value of $A_0$. The AFM used [28] is set into a glove box in which the PPM (part per million) of water is achieved, allowing the oscillator to keep a stable behavior over several weeks. Three surfaces were investigated, a silica surface, a silica grafted with aminopropyltriethoxysilanes molecules (APTES), and DNA deposited onto the grafted surface. APTES molecules were chosen because of their ability to stick DNA molecules onto surfaces [9, 10], while the silica surface is used as a reference. The experiments were performed with the same tip, without any evidence of change of its size and pollution. The stability of the tip was verified by recording time to time an approach-retract curve on the reference surface, the silica surface. The curve is performed at the oscillation amplitude showing the transition regime between intermittent contact situations and non-contact ones [21, 29]. This method is sensitive but does not mean that the tip has not been slightly modified. These measurements only indicate that if any modifications happen, there are weak enough to maintain the same attractive interaction between the tip and the surface.

Since the oscillation amplitude governs the strength of the attractive interaction, approach-retract curves were recorded at different working amplitudes $A_{free}$, ranging from 53 nm down to 4 nm. In order to minimize any asymmetric tip effect, the images presented in this work were all recorded by scanning along a direction perpendicular to the cantilever symmetry axis.

Silica and grafted silica surfaces were first thoroughly investigated; a detailed analysis is given in ref.[29]. Here are recalled a few results to help to understand the recorded image on DNA. For silica surface, the dominant repulsive regime is obtained down to the resonance amplitude $A_0 = 8$ nm. This low value indicates a rather small size of the tip apex. Those small tip sizes are scarcely available, typically, our own practice leads to the conclusion that over two hundred tips, no more than 10 percent do show such a low value. The APTES grafted silica surface exhibits a marked, stronger attractive interaction between the tip and the surface than the one of the silica alone.

Recording approach-retract curves [29] straightforwardly shows the observed difference between silica and grafted silica surfaces. Approach-retract curves give the variation of the oscillator's properties as a function of the tip-surface distance [21], thus are the required preliminary measurement to determine the experimental conditions to any recording image. A qualitative picture is readily obtained by looking at the amplitude at which the phase crosses the -90° value. Phase above (below) -90° indicates a dominant repulsive (attractive) regime. Figure 2 shows that, using $A_0 = 31$ nm, the recorded grafted surface image corresponds to a dominant attractive regime of the oscillator, while the silica surface image, in spite of being recorded at a smaller amplitude, keeps the oscillator within a repulsive regime. Because the tip's size is a constant, this result immediately indicates that the attractive interaction between the tip and the surface is much larger on the grafted surface than the one observed on the silica alone. The origin of the attractive interaction enhancement can be understood as follow : the pKa of APTES molecules is around 10 [9], thus for silica surface treated in aqueous solution at pH 3 to 7, amine groups can be chemically modified as ammonium species providing an additional attractive electrostatic contribution.

*II-2- Results on DNA*

The DNA height images shown in figures 3 and 4a were recorded using the repulsive regime (e.g. intermittent contact situations between the tip and the surface over the whole surface : $A_0 = 49$ nm and the average phase observed over the surface is -43°). To illustrate the main purpose of the present work, we focus on the images (height, left and phase, right) of the figure 4a. These images present the interest of showing a DNA molecule with its sequences nearly parallel either to the X or to the Y-axis. Therefore, a possible influence of an asymmetric shape of the tip, large enough to induce different strengths of the attractive interaction between the tip and the molecule, will be more easily seen since the tip crosses over the molecule along different directions.

The same molecule was observed at different driving amplitudes. Figure 4b shows the images (height and phase) obtained with $A_0 = 11$ nm. The height image shows a net difference. At high amplitude (fig.4a), the image exhibits a homogeneous contrast all along the molecule, while at driving amplitude $A_0 = 11$ nm, some parts of the molecule exhibit edges with a strong over-illumination. The driving amplitude $A_0 = 11$ nm and set point were chosen to image the grafted surface in a dominant attractive regime (the average phase of the surface is close to -123°). When this particular experimental condition is chosen, it occurs that above the DNA the oscillation behavior of the oscillator corresponds to a nearly dominant repulsive regime. The different regimes can be straightforwardly deduced from the recording of the approach-retract curves on the grafted surface, close to and above DNA. As stated above, the way the different regimes are discriminated is readily obtained by looking at the corresponding phase values (fig.5).

The over-illumination observed corresponds to an apparent higher height at proximity of the DNA, height which is not observed when the dominant repulsive regime is used over the whole surface (fig.4a, $A_0 = 49$ nm). This apparent increase of the height can be understood as follow. When the Tapping mode is used, the tip-sample distance is controlled through variations of the oscillation amplitude. Change of the oscillation amplitude is a consequence of change of the interaction between the tip and the surface. As shown in ref.[21], a decrease of the amplitude can either be due to repulsive or attractive interaction. When the attractive interaction between the tip and the surface becomes the leading one, the decrease of the amplitude is a function of the strength of the attractive interaction. Consequently, for non-contact situations, larger is the strength of the attractive interaction, larger is the closest distance between the tip and the surface at which a given reduction of the amplitude is reached. In other words, for the dominant attractive regime, a bright spot means that a higher attractive interaction is expected, provide that the surface does not show a bump structure at this particular location. Therefore, the image indicates an enhancement of the attractive interaction at proximity of the molecule, the molecule itself remaining imaged with the dominant repulsive regime (e.g. intermittent contact situations between the tip and the DNA).

The amplitude was further decreased in order to amplify the attractive interaction (figs.6). At $A_0 = 9$ nm (fig.6a), the apparent width of the molecule is widened. On the height image (left part on fig.6a), the few dark spots on the DNA correspond to a few remaining intermittent contact situations on the molecule. On the phase image (right part), the height dark spots become the bright ones since for these particular spots the phase is higher than the one given by non-contact situations. At the lowest amplitude, $A_0 = 6$ nm (fig.6b), the dominant attractive regime is obtained throughout the surface, including DNA. The molecule appears featureless and is further widened.

The marked evolution of the image as a function of the driving amplitude (figs.4a, 4b, 6a and 6b) can also be seen by reporting the corresponding cross sections (figs.7a, b, c and d respectively). At intermediate amplitude (fig.7b), $A_0 = 11$ nm, shoulders appear and the central part shows a small bump. The corresponding phase is above -90°. At $A_0 = 9$ nm, the small bump disappears but the shoulders remains quite well separated (fig.7c). At $A_0 = 6$nm, the shoulders widen and overlap leading to a featureless structure (fig.7d).

As shown with these sections, measured heights and widths are very much dependent of the experimental conditions chosen. This is not surprising since the oscillation behavior depends on the strength of the attractive interaction force. When the tip does not touch the surface or only slightly touches a soft material, there is no reason that the height variation uniquely gives a topographic information. In addition, the evolution of the cross sections does show that additional informations are obtained when the amplitude is reduced. This is particularly well verified at the intermediate amplitude ($A_0 = 11$ nm), where both the central part of the molecule and the shoulders are observed. As shown with the comparison given in figures 4, the contrast on the central part of the molecule is much better than the one obtained at the highest amplitude, showing unambiguously the superstructure of the molecule. In addition, the shoulders are quite narrow with a full width at half height around 5 nm, while at $A_0 = 9$ nm the full width is about 10 nm.

## III- Discussion

The use of different oscillation amplitudes reveals heterogeneous attractive interactions between the tip and the surface with a particular enhancement close to DNA that leads to a supplementary height of about half a nanometer. In addition, this enhancement of the attractive interaction is restricted to a limited lateral spatial extension of a few nanometers. Therefore, question rises about the physical origin of such an enhancement of the attractive interaction producing a very localized effect.

Since DNA molecule at pH 7 bears one negative charge per base, and that the grafted surface may also exhibits distribution of $NH_3^+$ species, a reasonable assumption is to consider an additional electrostatic interaction along the molecule. Nevertheless, considering a net electrostatic contribution of isolated charges gives a long-range interaction unable to describe the locality of the observed enhancement. The shortest range of attractive interaction is given by the Van der Waals dispersive interaction with an atom-atom interaction varying as $r^{-6}$, r being the distance between atoms. A similar short-range power law dependence is obtained when randomly oriented permanent dipolar momentums interact with a collection of non-polar species like the tip apex is. This additional Van der Waals like interaction is called Debye interaction [30]. If a short-range interaction is a necessary requirement, it is not enough to get a local interaction probed by the oscillating nanotip. For example, the usual Van der Waals sphere-plane interaction leads to an attractive interaction force between the tip and the surface varying as $\Delta^{-2}$, where $\Delta$ is the tip-surface distance. If random dipolar momentums are homogeneously distributed on a quite large area, the interaction with the tip will lead to a similar smooth variation as a function of the tip-sample distance. Thus, one cannot expect to get any drastic variation as a function of the (X;Y) location. By considering a line, or a narrow strip of randomly oriented dipoles, because the sum (the integration for the continuous case) is over a unique spatial dimension, the power law dependence, scaling as $\Delta^{-7/2}$ (see appendix), gives a more sensitive variation of the attractive interaction force as a function of the tip-sample distance, thus is able to provide a more localized effect of the attractive interaction.

The above assumption is supported by the possibility of having complexes made of negative and positive charges along the molecule. For instance, the presence of negative phosphate groups and positive amine groups are suitable candidates to built permanent dipoles. But other additional and significant contributions can also be issue either from the complexes made of negative phosphate groups and di-cations magnesium $M_g^{++}$ which have reacted together with the DNA molecules during the preparation of the solution or remaining water molecules [31]. Water molecules on DNA are known to be of importance for the location of di-cations [31] and because of their permanent dipoles, might also contribute to the enhancement of the attractive interaction. Note that, the contribution of a water meniscus in DFM experiments should have a less important effect than the one observed in AFM contact mode experiments [32], particularly for non-contact situations (low oscillation amplitudes) for which the tip does not touch the surface.

Whatever the precise chemical constitution of dipoles, let's consider that the distribution of randomly oriented dipoles forms a narrow strip on each side of the molecule. Using equ.7 of the appendix and considering two strips of dipoles, the numerical simulation (see appendix for more details) gives the cross sections reported in the figure 8a. When the oscillation amplitude is varied, the numerical results reproduce with a good agreement the general trends experimentally observed (fig.8a and 8b).

The results given by the simulation support the assumptions employed to interpret the over-illumination along the molecule. This result also suggests a mechanism describing the interaction between the DNA and the grafted substrate through a strong interaction between the APTES molecules and the phosphate groups. Here we face the difficult and long-standing problem of the substrate influence. For instance, the largest illumination, thus the highest attractive interaction, occurs above superstructures that are identified as plectonems [33-34] (fig.4). Such a superstructure is not expected for linearized molecules, except if a strong interaction due to the presence of cations or di-cations locally bends the molecule and screens the repulsive electrostatic interaction between phosphate groups [11, 33-35]. The structures observed together with the over-illumination suggest that the DNA molecule is tightly stuck onto the grafted substrate. Also, as for most of the numerous previous experiments [5, 6, 10], we do not get the expected DNA height of about 2 nanometers. As a matter of fact, except the noticeable result given in ref.[8], in which is clearly shown that the DNA molecule is loosely bounded on the substrate, most of the experimental results give DNA heights ranging between half and one nanometer. This deficit in height can be understood as another indication of a strong interaction between the molecule and the substrate.

**Acknowledgements**

It's a pleasure to thank E. Delain, D. Michel and E. Le Cam for fruitful discussions and for providing the DNA samples.

**Conclusion**

The present work was an attempt to image DNA molecule with a dynamic force microscope using various experimental conditions so that different properties of the DNA molecule deposited onto a substrate become accessible. The surface chosen is a silica surface grafted with silanes molecules ended with amine groups (APTES). Images were recorded at different oscillation amplitudes. By varying the oscillation amplitude, the attractive interaction between the oscillating tip and the DNA is varied. Use of small oscillation amplitudes favors the attractive interaction, while use of large amplitudes renders the contribution of the attractive interaction negligible. The images show a marked evolution as a function of the oscillation amplitude from which is deduced that the attractive interaction along the DNA is much larger than everywhere else on the surface. This enhancement of the attractive interaction is found to be amenable with a model considering a narrow strip of randomly oriented dipoles on each side of the molecule. The numerical simulations performed give a good agreement with the experimental results. The present work indicates that with the dynamic force microscopy, beyond studies of chain conformation, one can access to the distribution of interaction forces on the DNA molecules related to chemical properties.

**Appendix**

*A-1-   Attractive interaction between a sphere and a line of randomly oriented dipoles :*

The disperse part of the Van der Waals interaction between two identical atoms at a distance r is given by [30] :

$$V_{m-m}(D) = -\frac{C_{Disp}}{r^6},$$  (1)

with $C_{Disp} = 3\alpha_0^2 h\nu_I \big/ \left[4(4\pi\varepsilon_0)^2\right]$ the London parameter. $\alpha_0$ and $h\nu_I$ are the atom polarizability and ionization energy respectively.

The interaction between a non-polar molecule and a randomly oriented dipole (subscript "d") with a dipolar moment p at a distance r gives the same power law dependence :

$$V_{m-d}(D) = -\frac{C_{Debye}}{r^6},$$  (2)

with $C_{Debye} = 2\alpha_0 p^2 \big/ (4\pi\varepsilon_0)^2$ .

The attractive interaction between a sphere made of non-polar units and a randomly oriented dipole at a distance D is (fig.A1) [36] :

$$V_{m-s}(D) = -\frac{4\pi\rho R^3}{3} \frac{C_{Debye}}{(D+2R)^3 D^3},$$  (3)

with $\rho$ and R the density and radius of the sphere.

Now, to calculate the attractive interaction between a sphere and a line (subscript "l") of randomly oriented dipoles at a distance D with a linear density $\lambda$, the potential (3) has to be integrated all along that line :

$$V_{s-l}(D) = -2\int_0^\infty V_{m-s}(D_{eff}) \times \lambda dy,$$  (4)

where $D_{eff} = \sqrt{(R+D)^2 + y^2} - R$ . The calculation leads to :

$$V_{s-l}(D) = -\frac{\pi^2 \rho \lambda R^3 C_{Debye}}{2} \frac{1}{D^2(2R+D)^2 \sqrt{D^2 + 2RD}}$$  (5)

With Van der Waals interactions, the distance D at which the tip becomes sensitive to the surface is typically 1 nm : thus R/D>>1. This approximation allows equ.5 to be written as :

$$V_{s-l}(D) \overset{R \gg D}{\approx} -\frac{\pi^2 \rho \lambda \sqrt{R}}{8\sqrt{2}} \frac{C_{Debye}}{D^{5/2}} \qquad (6)$$

*A-2-    Numerical simulation of a Tapping mode cross section :*

The numerical simulations aimed to provide a support for describing the shoulders appearing at low amplitudes along the DNA molecule, e.g. in non-contact situation [36]. Two lines of randomly oriented dipoles are aligned at an arbitrary distance of 1 nm defining a "strip" of dipoles of 1 nm width. Two strips are considered separated by a distance of 7 nm. The geometry of the problem is sketched on fig.A1. The attractive interaction between the surface and the tip is described using a disperse Van der Waals interaction with a sphere-plane geometry.

The simulation is built in order to reproduce an experimental section and the feedback loop of the piezoelectric ceramic holding the sample. The set point at which the section is calculated is first reached with an approach curve. The approach is performed far from the location of the lines of dipoles ($X_{start}$ for instance on fig.A1). The vertical motion of the surface is stopped when the reduction of amplitude is equal to the one chosen.

During the scan[2], the program integrates the usual second kind differential equation of the forced, damped, oscillator plus two terms of forces due to the sphere-plane Van der Waals disperse interaction and to the sphere-strip of dipoles Debye interaction with a Runge-Kutta 4 method from arbitrary initial conditions :

$$\ddot{z}(t) + \frac{\omega_0}{Q}\dot{z}(t) + \omega_0^2 z(t) = \frac{F_{exc}}{m}\cos(\omega t) - \frac{HR}{6m[D_{sub} - z(t)]^2} - \frac{C'_{Debye}}{m}\sum_{n=1}^{4}\frac{1}{D_{dip(n)}(z(t);X(t))^{7/2}} \, , (7)$$

Amplitudes and phases of the oscillators are recorded thanks to a synchronic detection filtering the first harmonic. In equ.7, m, $\omega_0$ and Q are respectively the mass, resonance pulsation and quality factor of the oscillator. $F_{exc}$ and $\omega$ are the drive force and drive frequency defining the oscillations conditions. $D_{dip(n)}(z(t);X(t))$ is the distance between the tip and the $n^{th}$ line of dipoles defined as :

$$D_{dip(n)}(z(t);X(t)) = \sqrt{[R + z(t) - D_{sub}]^2 + [X_{dip(n)} - X(t)]^2} - R \qquad (8)$$

---

[2] The equation is identical for the approach curve except that the vertical distance is increased with time and X(t) = $X_{start}$, thus does not depend on the time.

The quality factor of the experimental oscillator is 410 (see text) and the frequency is around 153 kHz. Therefore the characteristic time of the oscillator is $2\times410/(153000)\approx5$ ms. This implies that the simulation sampling time is large enough in order to record oscillator's stationary state at each step of the calculation. This is validated by an adiabatic criterion calculating the number of points required for the whole scan.

**Captions**

Figure 1 :

a- Sketch of the dynamical system for two amplitudes A and A', with A>A'. The closest distance between the tip and the surface is a constant, e.g. D-A = D'-A'. D and D' are the distances between the surface and the equilibrium position at rest of the oscillator.

b- The force plot is calculated with a Van der Waals relationship between a sphere and a plane : $F_{vdW} = -\dfrac{HR}{6[D - z(t)]^2}$ , with $z(t) = A\cos(\omega t)$ the position of the tip as a function of time. H and R are respectively the Hamaker constant of the interface tip-surface and the radius of the tip. Their values have been fixed at $10^{-20}$ J and 10 nm. The values of A and A' are 49 and 11 nm (thin and thick continuous lines). D and D' were arbitrarily set to 49.5 and 11.5 nm. The plot illustrates that the lower the amplitude, the stronger the attractive force between the tip and the surface is.

Figure 2 :

Comparison between two approach-retract curves showing the variation of the phase as a function of the tip-sample distance on the silica (empty circles) and on the APTES surface (filled circles) with $A_0 = 21$ and 31 nm respectively. The experimental conditions are given in the text. The phase variation obtained on the silica indicates that the repulsive regime between the tip and the surface is dominant ($\varphi > -90°$) in spite of a smaller amplitude than on the APTES for which the attractive regime is dominant ($\varphi < -90°$). Thus the attractive interaction between the tip and the surface is much larger on the grafted surface than the one on the silica alone.

Figure 3 :

Height images of pSP65 DNA molecules observed on the grafted surface with $A_0 = 49$ nm.

Figure 4 :

a- Height (left) and phase (right) images of 200×200 $nm^2$ of the pSP65 molecule observed with $A_0 = 49$ nm. The vertical contrasts are 3 nm and 10°.

b- Same molecule, but with $A_0 = 11$ nm, thus in the attractive regime on the silanes. An over-illumination all along the edges of the molecule is clearly seen. The phase image shows that the DNA molecule is mostly imaged in the repulsive regime (see text and fig.5). The vertical contrasts are 3 nm and 90°.

Figure 5 :

Sections of the phase images of the figures 4a ($A_0 = 49$ nm) and 4b ($A_0 = 11$ nm) made on the central part of the DNA molecule. With $A_0 = 49$ nm, the repulsive regime is dominant over the whole surface, including DNA molecule. The average phase is –43° and the contrast is weak. This is the reason why the contrast of the image is 10°. With $A_0 = 11$ nm, the surface is mostly imaged with the attractive regime. The average phase value is around –123°. On the DNA, intermittent contact situation occurs with a phase value above –80°. This is the reason why the contrast of the image is 90°.

Figure 6 :

a- Identical zone than the one shown in figures 4, but with $A_0 = 9$ nm. The apparent width of the molecule is larger than the one at higher amplitudes. The vertical contrasts are 3 nm and 10°.

b- Same zone, but with $A_0 = 6$ nm. The attractive regime is dominant all over the area. The molecule appears featureless. The vertical contrasts are 3 nm and 10°.

Figure 7 :

Sections of the height images of the figures 4a, 4b, 6a and 6b made on the central zone of the molecule. Reducing the amplitude from $A_0 = 49$ (a) down to 6 nm (d) leads to an increase of the attractive interaction between the tip and the surface. As a consequence, the shoulders appearing on the edges of the molecule with $A_0 = 11$ nm (b) become to overlap.

Figure 8 :

a- Numerical simulations of a section with two strips of dipoles obtained in the Tapping mode with two amplitudes. The strips of dipoles are 1 nm width and localized at X = +3.5 and –3.5 nm. The numerical parameters are p = 30 D with a linear density of 10 $nm^{-1}$ (see appendix). The numerical oscillations conditions are $A_0 = 15$ (thin line) and 7 nm (thick line).

b- Zoom of the experimental cross sections of the figures 7c and 7d corresponding to the amplitudes $A_0 = 9$ and 6 nm.

# Figures

Figure 1:

**a:**

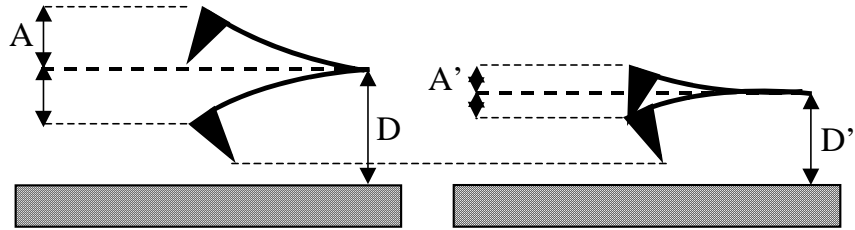

**b:**

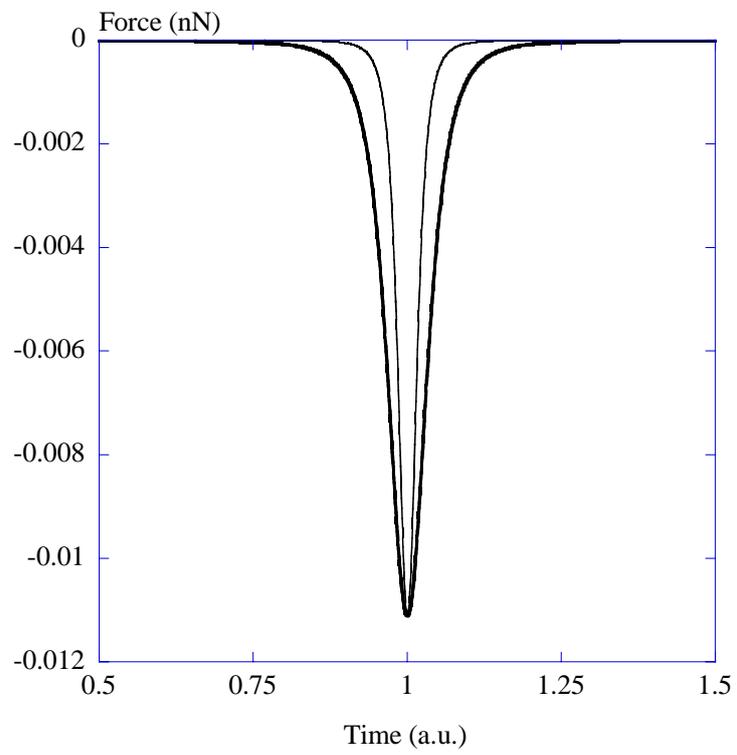

Figure 2:

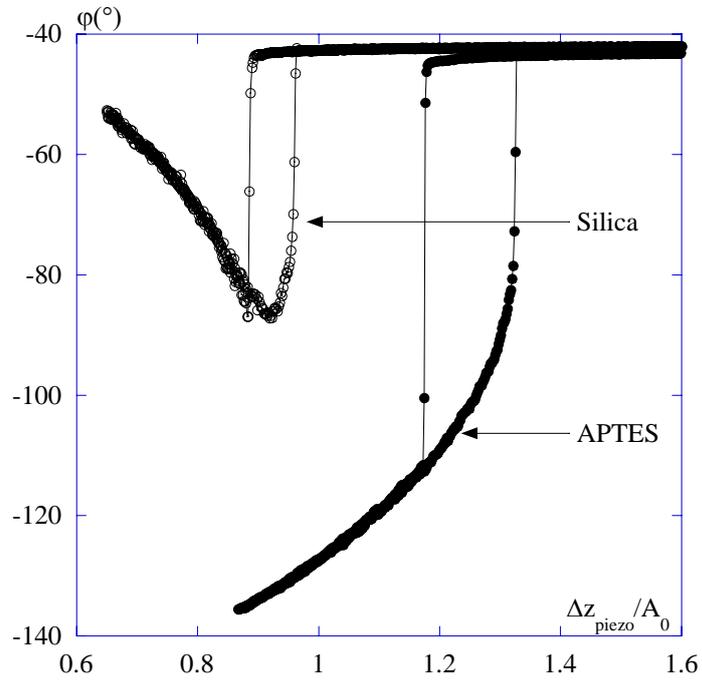

Figure 3:

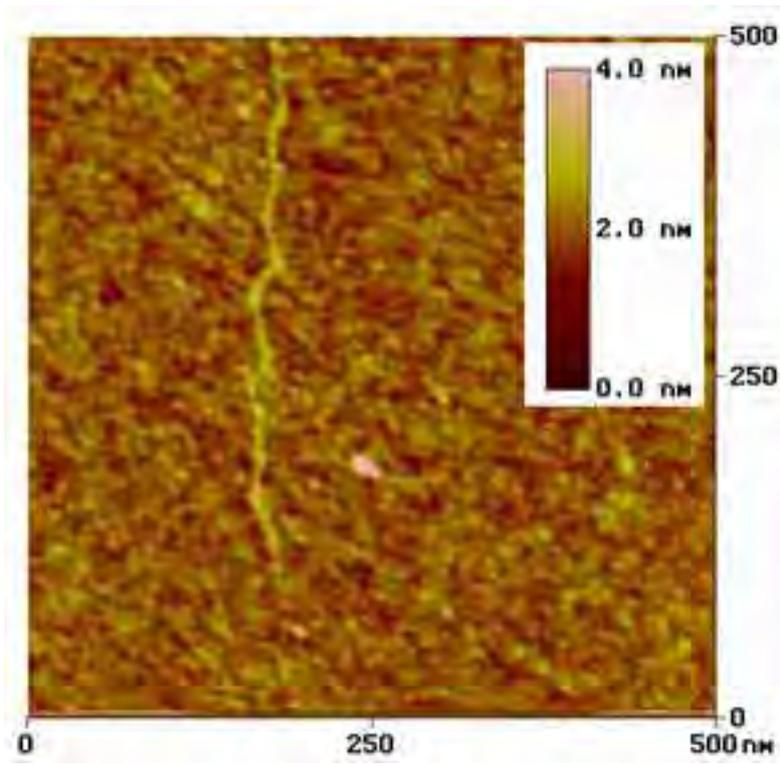

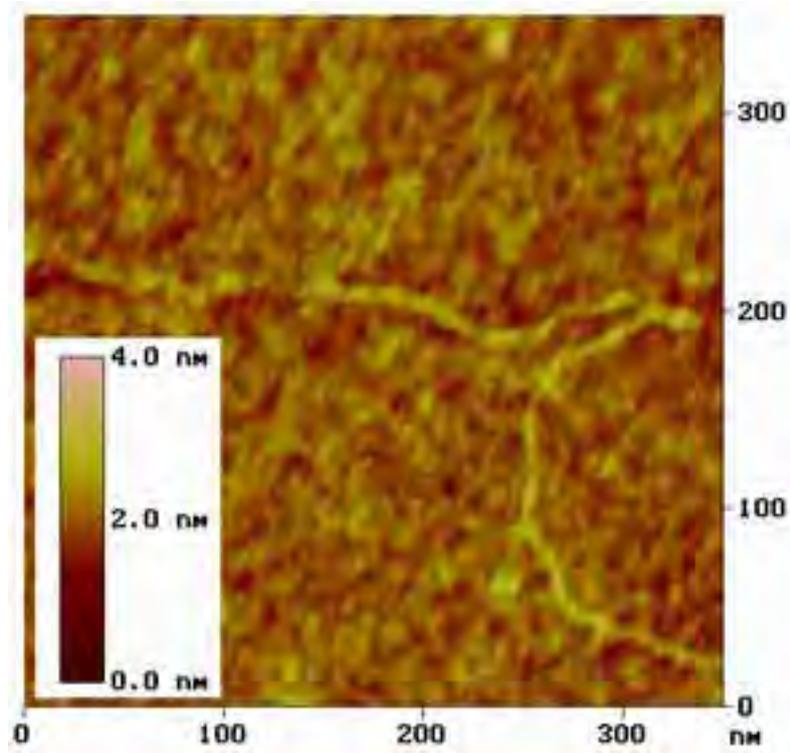

Figure 4:

**a:**

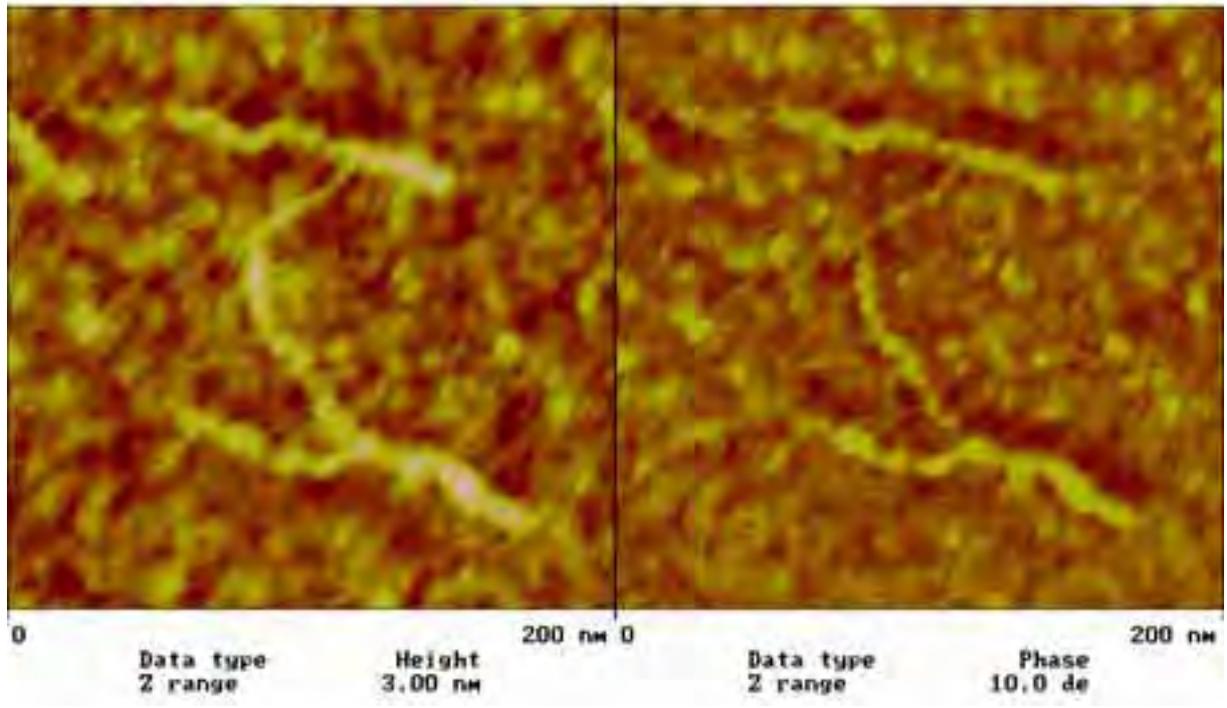

**b:**

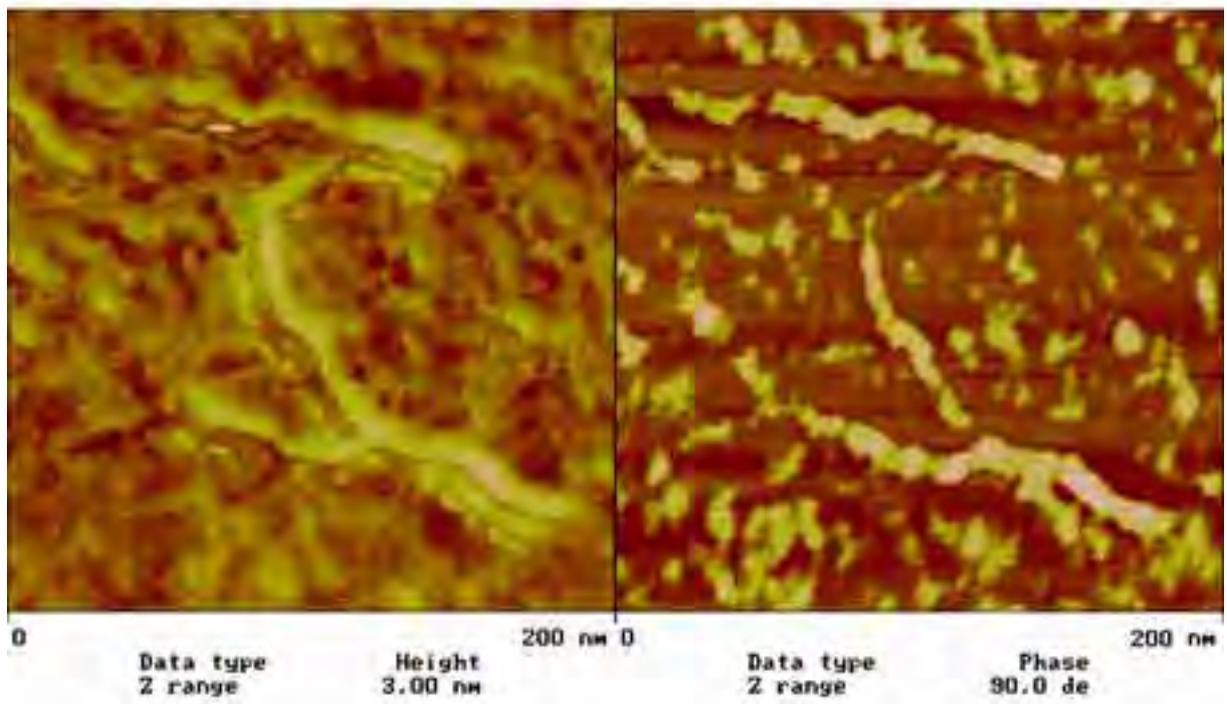

Figure 5:

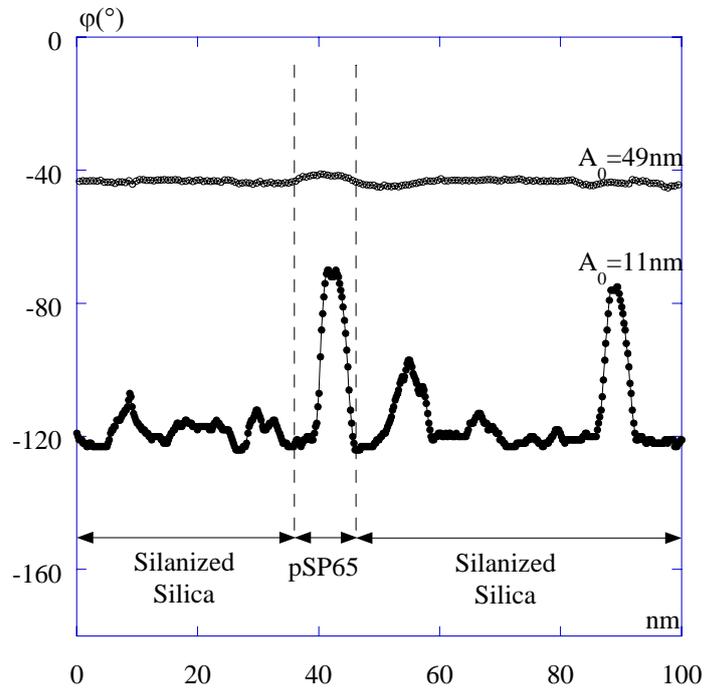

Figure 6:

**a:**

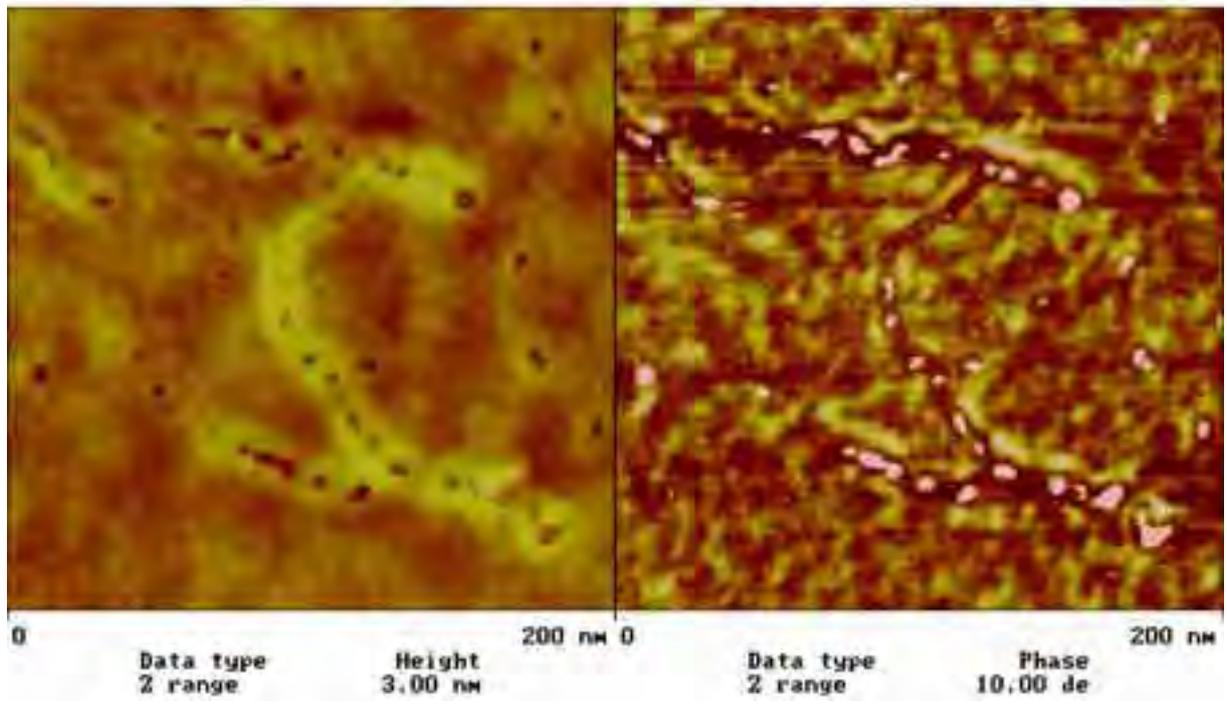

**b:**

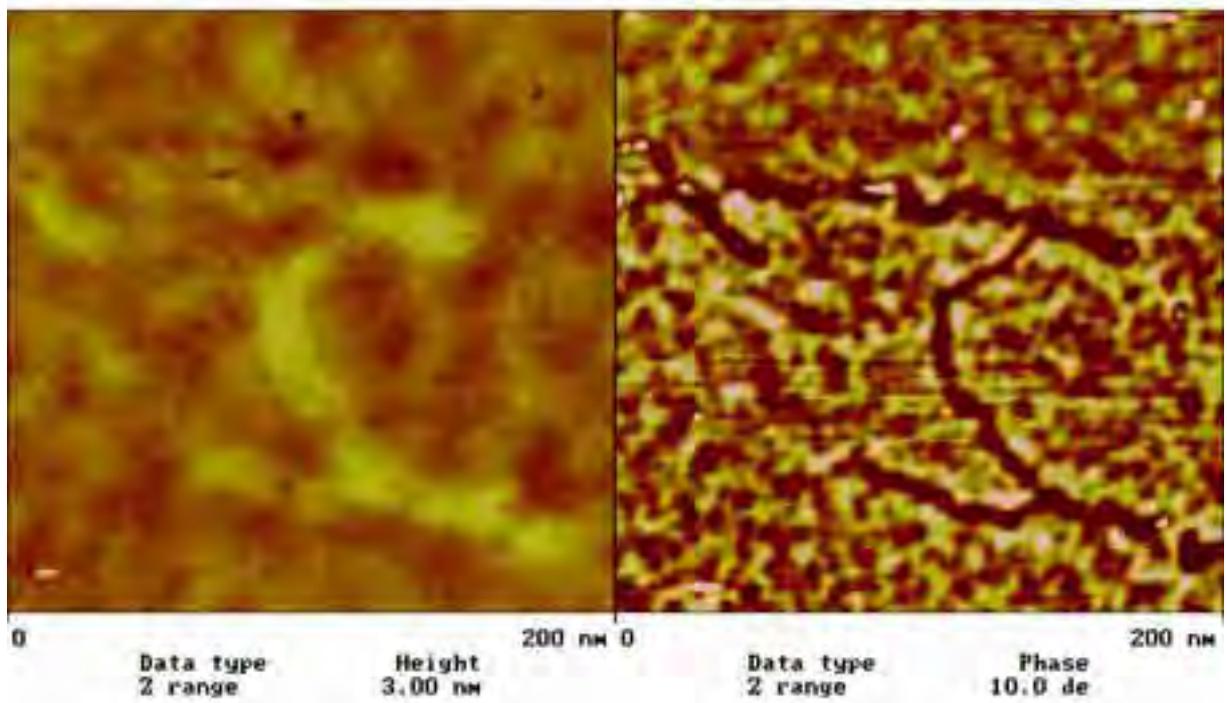

Figure 7:

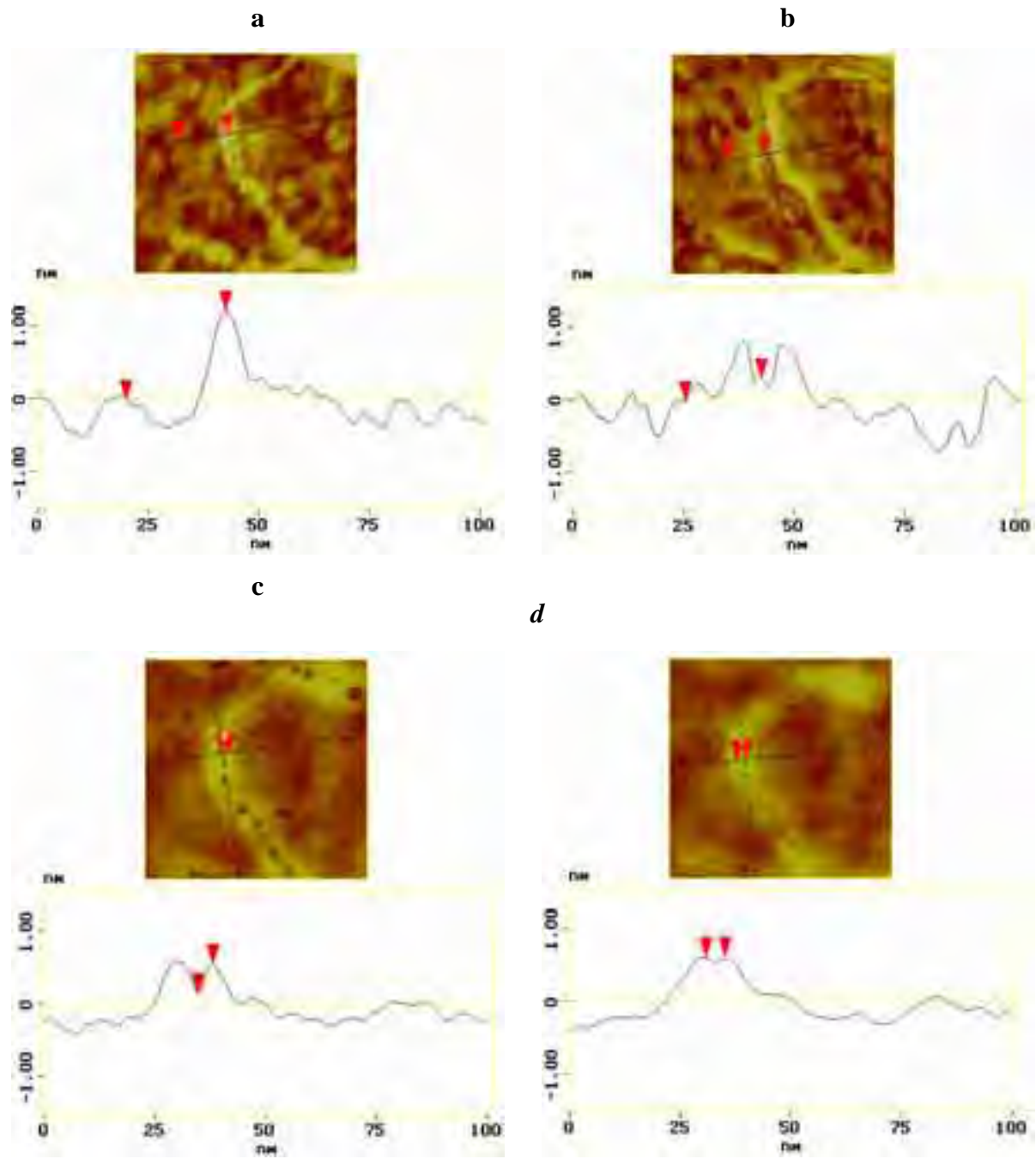

Figure 8:

**a:**

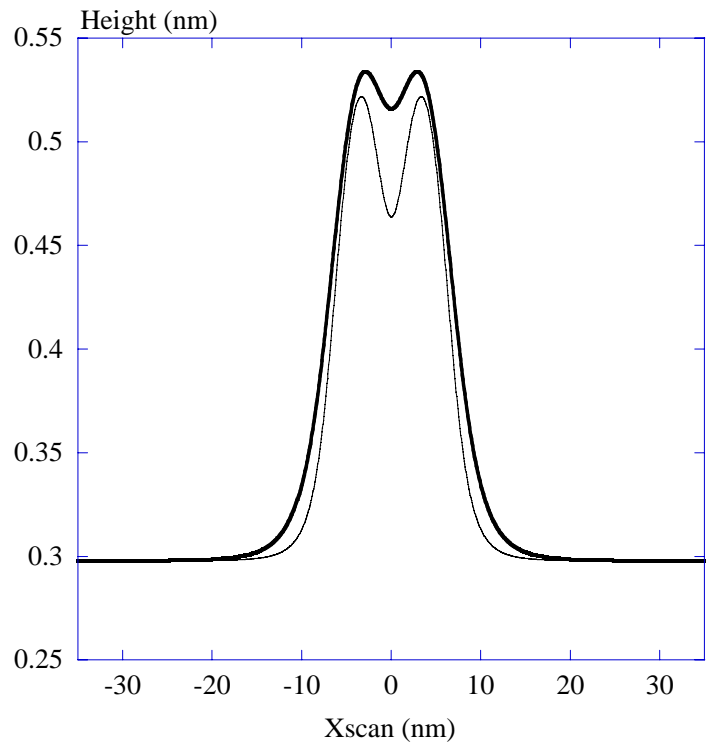

**b:**

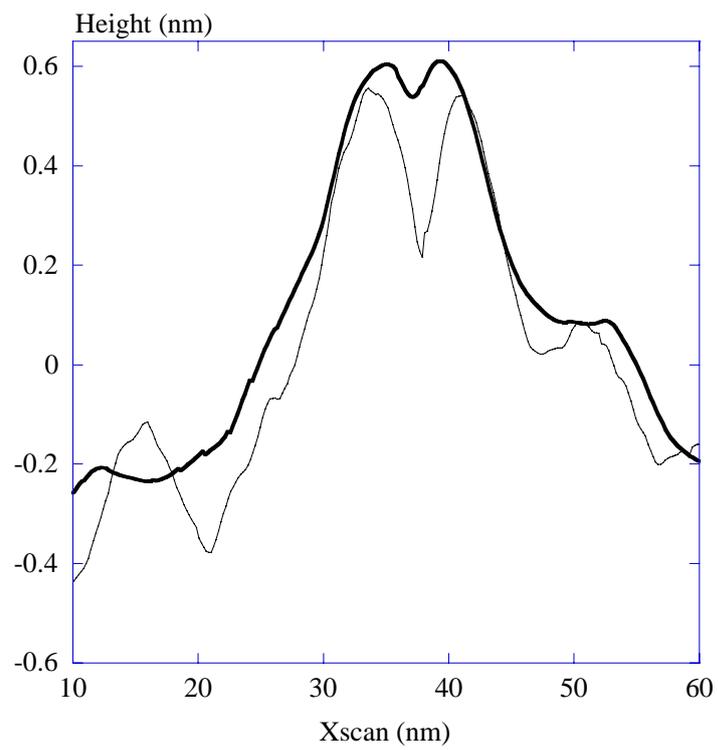

Figure A1 :

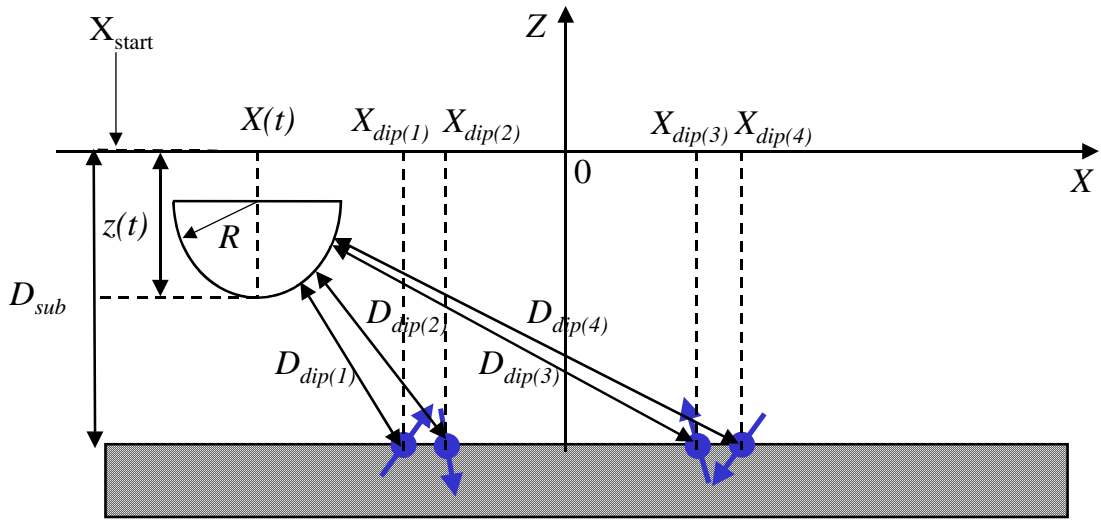